\documentclass[aps,prl,amsmath,amssymb,citeautoscript,reprint]{revtex4-2}
\usepackage{amsmath, amssymb, amsthm}
\usepackage{graphicx, float, xcolor, pgf}
\usepackage{physics, siunitx}
\usepackage[caption=false, justification=centerlast]{subfig}
\let\autocite\cite
\makeatletter
\let\oldtheequation\theequation
\renewcommand\tagform@[1]{\maketag@@@{\ignorespaces#1\unskip\@@italiccorr}}
\renewcommand\theequation{(\oldtheequation)}
\makeatother

\usepackage{hyperref}
\hypersetup{
  pdfauthor={Amartya Bose},
  pdftitle={Impact of Spatial Inhomogeneity on Excitation Energy Transport in the Fenna-Matthews-Olson Complex},
  colorlinks=true,
  allcolors=.
}

\begin{document}
\title{Impact of Spatial Inhomogeneity on Excitation Energy Transport in the Fenna-Matthews-Olson Complex}
\author{Amartya Bose}
\email{amartya.bose@tifr.res.in}
\email{Both authors contributed equally to this work.}
\affiliation{Department of Chemical Sciences, Tata Institute of Fundamental Research, Mumbai 400005, India}
% \affiliation{Department of Chemistry, Princeton University, Princeton, New Jersey 08544, USA}
\author{Peter L. Walters}
\email{peter.l.walters2@gmail.com}
\email{Both authors contributed equally to this work.}
\affiliation{Department of Chemistry, University of California, Berkeley, California 94720, USA}
\affiliation{Miller Institute for Basic Research in Science, University of
    California Berkeley, Berkeley, California 94720, USA}
\allowdisplaybreaks

\begin{abstract}
    The dynamics of the excitation energy transfer (EET) in photosynthetic
    complexes is an interesting question both from the perspective of
    fundamental understanding and the research in artificial photosynthesis.
    Challenges persist in numerically simulating these systems both in
    parameterizing them and following their dynamics over long periods of time.
    Over the past decade, very accurate spectral densities have been developed
    to capture spatial inhomogeneties in the Fenna-Matthews-Olson (FMO) complex.
    We investigate the dynamics of FMO with an exact treatment of various
    theoretical spectral densities. Because FMO has Hamiltonian elements that connect most
    of the bacteriochlorophyll sites together, it becomes difficult to
    rigorously identify the energy transport pathways in the complex. We use the
    recently introduced ideas of relating coherence to population derivatives to
    analyze the transport process and reveal some of the pathways.
\end{abstract}
\maketitle

Light harvesting complexes (LHCs) play an important role in photosynthesis in a
host of plants, bacteria and algae. The so-called ``antenna complexes'' capture
solar energy, converting it into an electronic excitation, and carries it to the
reaction center where charge separation leads to further chemistry. Crucially,
these systems form the basis and inspiration for attempts at artificial
photosynthesis. Thus, understanding the mechanisms that allow for the efficient
transport of the molecular excitation from the point of creation to the reaction center is of
fundamental importance. A lot of work has been done in simulating the excitation
energy transport (EET) and characterizing the vibronic
couplings~\cite{rengerTheoryExcitationEnergy2009,
    caoQuantumBiologyRevisited,saitoSitedependentFluctuationsOptimize2019,
    cignoniAtomisticModelingLightharvesting2022, maityRecentProgressAtomistic2022,
    cuiRolePigmentProtein2021}. Early
experiments~\autocite{savikhinOscillatingAnisotropiesBacteriochlorophyll1997,
    engelEvidenceWavelikeEnergy2007} seemed to provide evidence of quantum beating.
Theoretical studies were performed around the same time to shed light on the
origins of these long-lived electronic
oscillations~\autocite{ishizakiAdequacyRedfieldEquation2009,
    ishizakiUnifiedTreatmentQuantum2009, ishizakiTheoreticalExaminationQuantum2009,
    sarovarQuantumEntanglementPhotosynthetic2010}. It was hypothesized that this
oscillatory dynamics could be the reason behind the efficiency of EET in
biosystems. However, more recent experimental
investigations~\autocite{duanNatureDoesNot2017,
    thyrhaugIdentificationCharacterizationDiverse2018} have shown that the optical
2D photon echo spectra at ambient temperature does not show long-lived
electronic quantum coherence.

Theoretical studies have been widely performed using
Redfield~\autocite{ishizakiAdequacyRedfieldEquation2009} and F{\"o}rster
theory~\cite{forsterZwischenmolekulareEnergiewanderungUnd1948}. However, the
applicability of these approximate perturbative techniques is not always
guaranteed \textit{a priori}. Simulations of thermal dynamics at the ambient
temperature is optimally performed using reduced density matrix-based approaches
like the hierarchical equations of
motion~\autocite{tanimuraTimeEvolutionQuantum1989,
    tanimuraNumericallyExactApproach2020} (HEOM) or path integrals using the
Feynman-Vernon influence functional~\autocite{feynmanTheoryGeneralQuantum1963}.
While the quasi-adiabatic propagator path
integral~\cite{makriTensorPropagatorIterativeI1995,makriTensorPropagatorIterativeII1995}
method (QuAPI) has been used to study the Fenna-Matthews-Olson (FMO)
complex~\autocite{nalbachRoleDiscreteMolecular2012}, almost every other
contemporary and more recent studies of the exact dynamics seem to have focused
on using HEOM as the method of
choice~\autocite{ishizakiTheoreticalExaminationQuantum2009,
    strumpferEffectCorrelatedBath2011, strumpferExcitedStateDynamics2012,
    wilkinsWhyQuantumCoherence2015, cuiRolePigmentProtein2021}. EET problems have
also been studied using semiclassical
methods~\autocite{taoSemiclassicalDescriptionElectronic2010,
    nalbachIterativePathintegralAlgorithm2011, leeSemiclassicalPathIntegral2016,
    leeModelingElectronicNuclearInteractions2016,
    mulvihillSimulatingEnergyTransfer2021}. Advances in path integral-based
methods~\cite{makriSmallMatrixDisentanglement2020, makriSmallMatrixPath2020}
have recently made them lucrative for studying EET
systems~\cite{kunduRealTimePathIntegral2020, boseAllModeQuantumClassical2020,
    boseTensorNetworkPath2022, kunduB800toB850RelaxationExcitation2022,
    kunduTightInnerRing2022}. Among these advances, there have been developments
that combine ideas from tensor network and influence functional to help
alleviate the cost of path integral calculations in different
ways~\cite{strathearnEfficientNonMarkovianQuantum2018,jorgensenExploitingCausalTensor2019,
    boseTensorNetworkRepresentation2021, bosePairwiseConnectedTensor2022,
    boseMultisiteDecompositionTensor2022, boseEffectTemperatureGradient2022,
    boseTensorNetworkPath2022}.

\begin{figure*}
    \subfloat[FMO trimer with protein
        scaffold]{\includegraphics[scale=0.20]{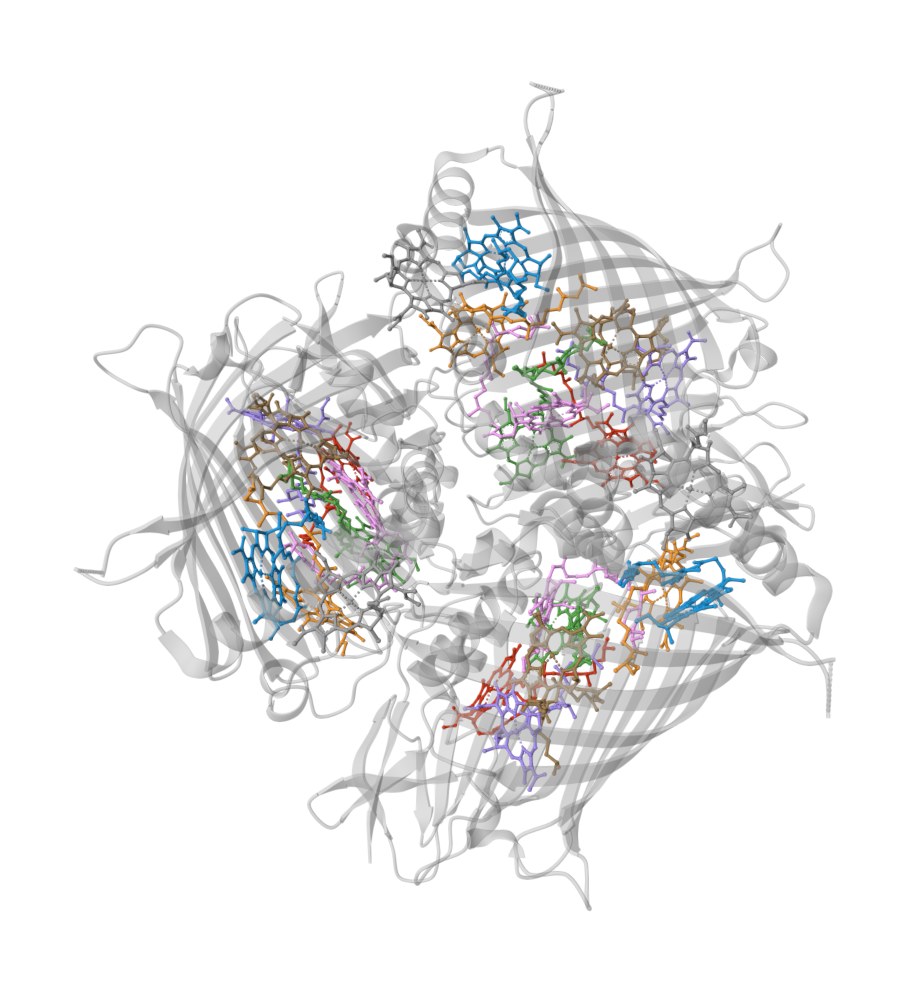}}
    ~\subfloat[FMO monomer with bacteriochlorophyll molecules
        labelled.]{\includegraphics[scale=0.35]{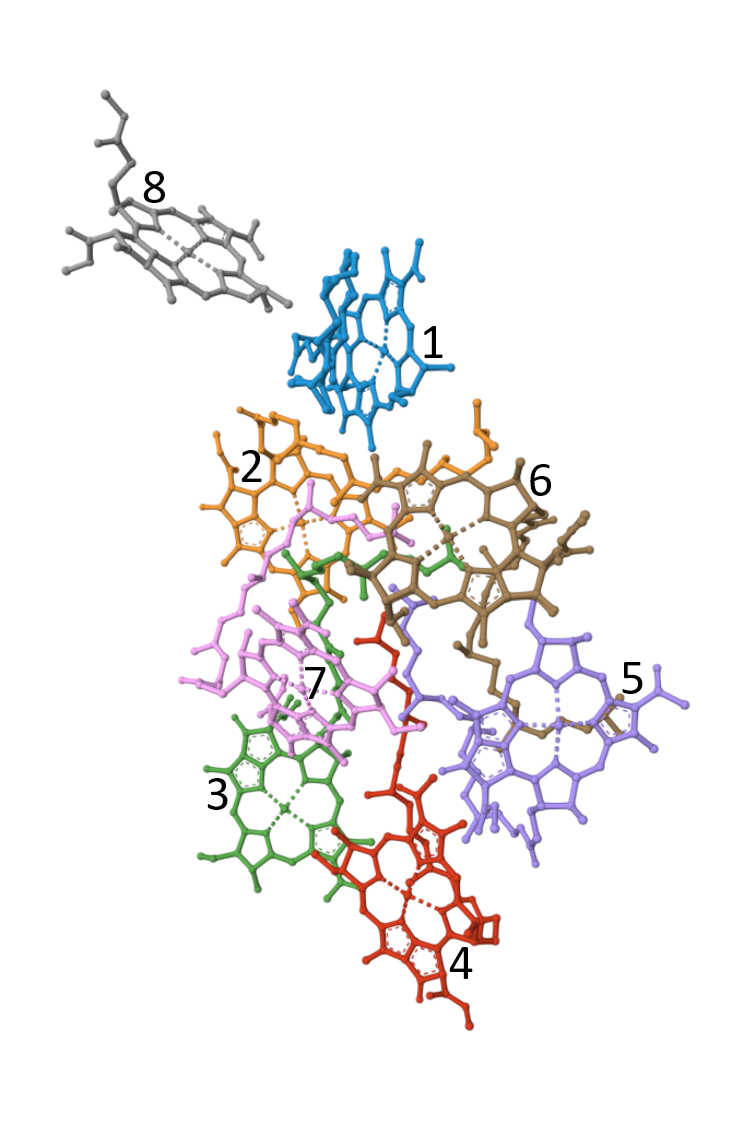}}
    \caption{Fenna-Matthews-Olson Complex in its trimeric and monomeric forms.
        Blue: BChl 1. Orange: BChl 2. Green: BChl 3. Red: BChl 4. Purple: BChl 5.
        Brown: BChl 6. Pink: BChl 7. Gray: BChl 8.}\label{fig:FMO}
\end{figure*}

In this work, we simulate the EET dynamics in the FMO complex with accurate
descriptions of the vibrational degrees of freedom. The usual practice is to
focus the numerical studies on the evolution of the population in the excited
states of each of the chromophores. Here, we additionally use the recently
developed coherence-based analytic
techniques~\cite{wuEfficientEnergyTransfer2012, daniQuantumStatetoStateRates2022,
    daniTimeEvolvingQuantumSuperpositions2022,
    boseImpactSolventStatetostate2023} to unravel the pathways that the
excitation takes and the impact of the vibrational modes.
\citet{bakerRobustnessEfficiencyOptimality2015} have explored these pathways in
FMO using the Lindblad master equation. Their approach based on knocking
particular chromophores out of the system and evaluating the resulting change in the
transport properties. Thus, they were able to ascertain the importance of particular
chromophores to the transport process.
% \textcolor{blue}{\citet{saitoSitedependentFluctuationsOptimize2019} have
%     recently demonstrated the importance of site-dependent fluctuations in FMO using
%     Drude-Lorentz spectral densities that approximate the atomistically computed
%     ones.}

The methods that simulate the time evolution of the reduced density matrix,
typically, require the calculation of spectral densities to characterize the
effect of the protein scaffolding and the localized vibrations of the
chlorophyll molecules. The spectral density can be calculated as the Fourier
transform of the energy gap correlation
function~\cite{makriLinearResponseApproximation1999,
    boseZerocostCorrectionsInfluence2022}. In case of an EET, this is equivalent to
the autocorrelation function of excitation energy fluctuations of individual
chromophores. While a lot of work has been done trying to characterize the
site-dependent spectral densities for the
FMO~\cite{klingerNormalModeAnalysis2020,
    olbrichTheorySimulationEnvironmental2011, rengerNormalModeAnalysis2012,
    chailletStaticDisorderExcitation2020}, there seem to be significant
disagreements on the best way of obtaining them. Zerner's intermediate neglect of
differential orbital method (ZINDO) and time-dependent density functional theory
(TD-DFT) have been used to calculate the excited state energy fluctuations along
classical molecular dynamics (MD) trajectories run on the ground
Born-Oppenheimer (BO)
surface~\autocite{olbrichTheorySimulationEnvironmental2011,
    olbrichTimeDependentAtomisticView2010, maityDFTBMMMolecular2020}. Extensive
analysis has been done over the past decade on how to best and most consistently
capture the fine interplay of dynamics and electronic structure that goes into
these spectral densities. Because of the so-called ``geometry mismatch'' problem
stemming from the inconsistencies between the conformations obtained from these
purely classical MD trajectories and the subsequent quantum mechanical
calculations, mixed QM/MM dynamics simulations on the ground BO surface have also been
performed~\autocite{blauLocalProteinSolvation2018, maityDFTBMMMolecular2020}.

Under physiological conditions, FMO exists as a trimer with each monomer
consisting of eight bacteriochlorophyll-a (BChl a) units. The structure of the
complex is shown in Fig.~\ref{fig:FMO}. The Hamiltonian describing the
EET process in one 8-unit monomer can be expressed by a Frenkel model,
\begin{align}
    \hat{H}_0 & = \sum_{j=1}^8 \epsilon_j\dyad{j} + \sum_{k\ne j} h_{k,j} \dyad{j}{k},
\end{align}
where $\epsilon_j$ is electronic excitation energy of the $j$th BChl molecule in the absence of
the protein environment, and $h_{k,j}$ represents the electronic couplings. The
state where only BChl $j$ is excited is denoted by $\ket{j} =  \ket{e_j} \otimes \prod_{k\ne j}
    \ket{g_k}$. Here, $\ket{g_j}$ and $\ket{e_j}$ represents the
local ground and excited states of the $j$th BChl unit. The full dissipative
environment including contributions from both the local, rigid vibrations and
the dynamical environment of the protein scaffolding is characterized by a
harmonic bath on each site, $j$:
\begin{align}
    \hat{H}^j_B & = \sum_\xi \frac{p_{j\xi}^2}{2 m_{j\xi}} + \frac{1}{2} m_{j\xi} \omega^2_{j\xi} \left(x_{j\xi} - \frac{c_{j\xi} \hat{s}_j}{m_{j\xi} \omega_{j\xi}^2}\right)^2,\label{eq:sys_bath}
\end{align}
where $\omega_{j\xi}$ and $c_{j\xi}$ are the frequency and coupling of the
$\xi$th mode on BChl $j$. The bath interacts with the $j$th BChl unit through
the diagonal operator $\hat{s}_j$ specified by $\hat{s}_j\ket{g_j} = 0$ and
$\hat{s}_j\ket{e_j}=\ket{e_j}$. The frequencies and couplings of the bath are
related to the spectral density as follows:
\begin{align}
    J_j(\omega) & = \frac{\pi}{2}\sum_\xi \frac{c_{j\xi}^2}{m_{j\xi}\omega_{j\xi}}\delta(\omega_{j\xi}-\omega).
\end{align}

Thus, the full universe including the system and the bath is defined by the
following Hamiltonian, which has a Frenkel-Holstein structure:
\begin{align}
    \hat{H} & = \hat{H}_0 + \sum_{j=1}^8 \hat{H}^j_B.\label{eq:system_bath}
\end{align}
In many cases, the system Hamiltonian, $\hat{H}_0$, is specified in terms of the
site energy (a.k.a the optical excitation energy), $E_j$. This optical
excitation energy is however dependent upon the environment. Therefore,
$\epsilon_j$ is obtained after shifting the optical excitation energy by the
corresponding reorganization energy~\cite{ishizakiUnifiedTreatmentQuantum2009}.
%Further details is provided in the Appendix.

Under the influence of the thermal vibrational baths, the reduced density
matrix corresponding to the EET system at time $t=N\Delta t$ is given by a
path integral expression,
\begin{align}
    % \mel{S_N^+}{\tilde\rho(N\Delta t)}{S_N^-} & = \sum_{\{S_j^\pm\}} P^{(0)}_{S^\pm_0, S^\pm_1\ldots S^\pm_N} F[S_0^\pm, S^\pm_1\ldots S^\pm_N] \\
    % F[S_0^\pm, S^\pm_1\ldots S^\pm_N]         & = \prod_{l=1}^8 \exp\left(-\sum_k(s_{k,l}^+-s_{k,l}^-)\right.\nonumber                          \\
    %                                           & \times \left.\sum_{k'\le k}(\eta_{kk'}^{(l)}s^+_{k',l} - \eta_{kk'}^{(l)*}s^-_{k',l})\right)
    \tilde\rho(S^\pm_N, N\Delta t) & = \sum_{S_0^\pm}\sum_{S^\pm_1}\cdots\sum_{S^\pm_{N-1}} \tilde\rho(S^\pm_0, 0)P_{S^\pm_0, S^\pm_1\ldots S^\pm_N}                                  \\
                                   & = \sum_{S_0^\pm}\sum_{S^\pm_1}\cdots\sum_{S^\pm_{N-1}} \tilde\rho(S^\pm_0,0) P^{(0)}_{S^\pm_0, S^\pm_1\ldots S^\pm_N} F[\left\{S^\pm_n\right\}],
\end{align}
where $P_{S^\pm_0, S^\pm_1, \ldots S^\pm_N}$ is the path amplitude tensor which
represents the amplitude of the system for moving along the specified sequence of
forward-backward states in presence of the solvent. In the
notation used here, $S^\pm_n$ represents the collective forward-backward state of
the system at the $n$th time point. (I.e., $S^\pm_n =
    \left\{s^\pm_{1,n},s^\pm_{2,n}\ldots s^\pm_{8,n}\right\}$, where $s^\pm_{j,n}$
is the forward-backward state of the $j$\textsuperscript{th} site at the
$n$\textsuperscript{th} time point.) The path amplitude tensor is a product of
the ``bare'' path amplitude tensor, $P^{(0)}_{S^\pm_0, S^\pm_1\ldots S^\pm_N}$,
representing the amplitude of the isolated system for moving along the same
points and the Feynman-Vernon influence
functional~\cite{feynmanTheoryGeneralQuantum1963}, $F$, representing the impact
of the solvent degrees of freedom on the system. The dynamics of the isolated
system is Markovian. It is the presence of the influence functional, $F$, that
induces non-Markovianness in the dynamics. While formally this expression
depends on the entire history of any path going back to time step 0, in
condensed phases the memory dies out and is calculated only till a finite number
of time-steps, $L$. This $L$ is a convergence parameter.

% where $P^{(0)}_{S^\pm_0, S^\pm_1\ldots S^\pm_N}$ is the bare path amplitude
% tensor which gives the amplitude of the isolated system of moving along a particular
% forward-backward path encoded by $S^\pm_0, S^\pm_1, \ldots, S^\pm_N$. At any
% time-point $j$, $S^\pm_j$ is a short-hand for $S^+_j$ and $S^-_j$ which
% represents the forward and backward states of the system at that time-point. For
% the case of Eq.~\ref{eq:system_bath}, for every $j$, $S_j^+$ and $S_j^-$ refer
% to a vector containing the expectation values of the interaction operator
% $\dyad{j}$, each of which are represented by $s_{j,l}$ with $l=1\ldots 8$
% representing the state of the $l$th BChl unit. The propagation encoded in
% $P^{(0)}$ is fully Markovian. The ``influence'' of the baths on the system
% modulates this bare amplitude through $F$ which is the temporally non-local
% Feynman-Vernon influence functional~\cite{feynmanTheoryGeneralQuantum1963}. The
% $\eta_{kk'}^{(l)}$ coefficients are the discretized bath response functions for
% the $l$th monomer which is responsible for causing the non-Markovian
% dynamics~\cite{makriTensorPropagatorIterativeII1995}. The length of this
% non-locality is a measure of the non-Markovian nature of the problem. Typically,
% in condensed phases, the non-Markovianness can be truncated at a finite length
% of history, which is used as a convergence parameter.

\begin{figure}
    \subfloat[Path amplitude tensor as an MPS.]{\includegraphics[scale=0.30]{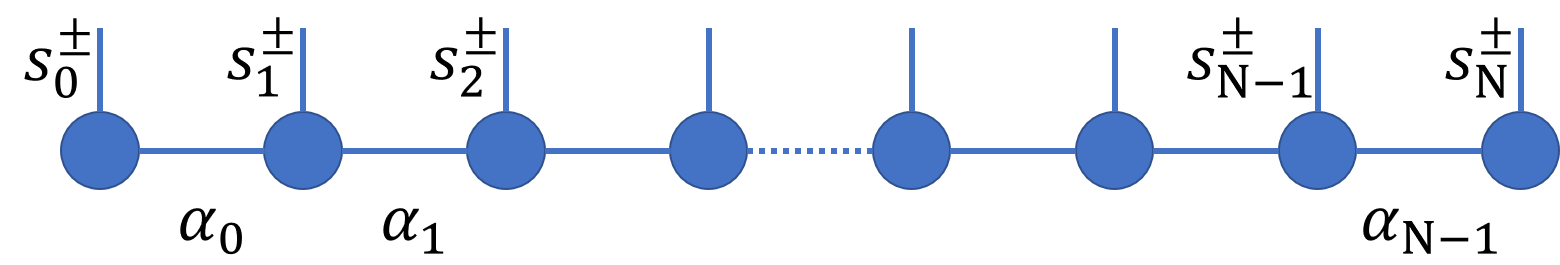}}
    
    \subfloat[Feynman-Vernon influence functional as an MPO.]{\includegraphics[scale=0.30]{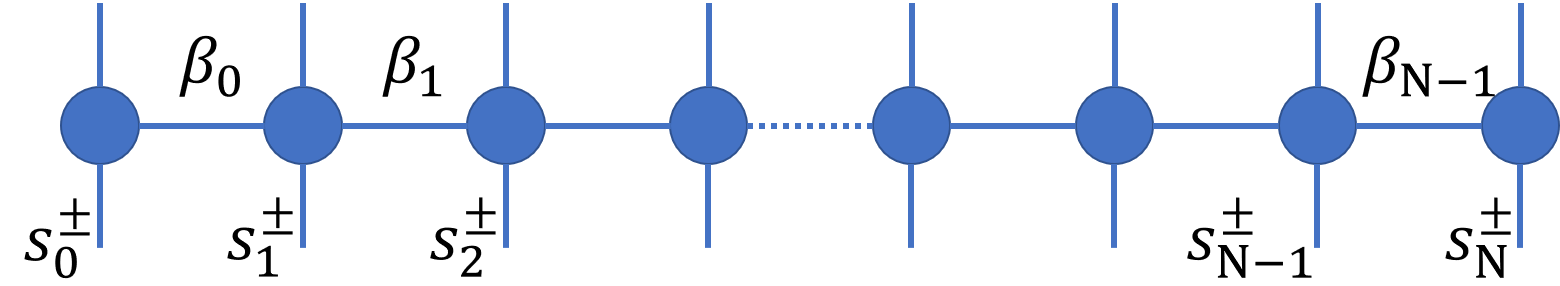}}
    \caption{Matrix product representations of the path amplitude tensor and the influence functional operators.}\label{fig:MPS_MPO}
\end{figure}
The dimensionality of the path amplitude tensor grows as $d^{2L}$ where the
system is $d$-dimensional (in this case, $d=8$), and $L$ is the
memory length. In the time-evolving matrix product operators
(TEMPO)~\cite{strathearnEfficientNonMarkovianQuantum2018} method, this exponential
growth is heavily compressed by recognizing that the correlations between
points separated by large time-spans becomes negligible even in the presence of
a non-Markovian bath. This suggests that one can use a matrix product state
(MPS) to efficiently represent the path amplitude tensor. An MPS or a tensor
train form is obtained by performing a sequential truncates singular value
decompositions, so that each of the indices on the original tensor can be
attributed to separate low-ranked tensors. In this representation, the terminal
tensors are rank-2 and the intermediate tensors are rank-3 as shown in
Fig.~\ref{fig:MPS_MPO}~(a). The common index between any two neighboring tensors
and its dimension are called the bond index and bond dimension respectively. In
such a representation, it can be shown that the influence functional can
similarly be analytically written as a matrix product operator
(MPO)~\cite{boseTensorNetworkRepresentation2021}, which is shown in
Fig.~\ref{fig:MPS_MPO}~(b). This analytic form is optimizes the representation
accounting for the multiple baths and the symmetries in the influence functional
expression~\cite{boseTensorNetworkRepresentation2021}. While the bare path
amplitude has zero long-distance correlations, and consequently a very compact
MPS representation, subsequent applications of the influence functional MPO
builds up these correlations, leading to an increase in the bond dimension.
However, efficient algorithms exist to minimize the growth of the bond dimension
of the MPS on application of an MPO based on convergence
parameters~\cite{paeckelTimeevolutionMethodsMatrixproduct2019}. This combined
with an optimal representation of the influence functional MPO makes the tensor
network approach especially efficient at simulating these systems. This method
is used for all the simulations in this paper.
% While it is well-known that
% for independent baths coupled in this manner the
% influence functional becomes a product of the individual influence functionals
% associated with each bath, TNPI enables an analytic representation of this
% product into a single matrix product operator (MPO). This makes simulations
% significantly more efficient.
% \begin{figure}
%     \centering
%     \subfloat[Comparison with classical MD spectral density from Ref.~\cite{olbrichTheorySimulationEnvironmental2011}]{\includegraphics{fig3a.pdf}}
% 
%     \subfloat[Comparison with experiment]{\includegraphics{fig3b.pdf}}
%     \caption{Comparison of spectral densities averaged across all the eight BChl molecules.}\label{fig:avg_spectral_densities}
% \end{figure}
\begin{figure}
    \centering
    \subfloat[Comparison with ZINDO-based classical MD spectral density from Ref.~\cite{olbrichTheorySimulationEnvironmental2011}]{\includegraphics{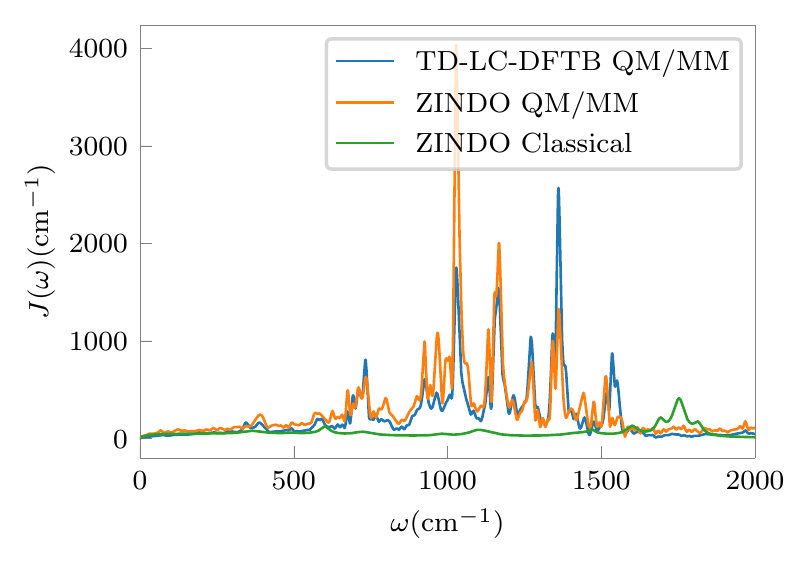}}
    
    \subfloat[Comparison with experiment~\cite{ratsepElectronPhononVibronic2007, kellShapePhononSpectral2013}]{\includegraphics{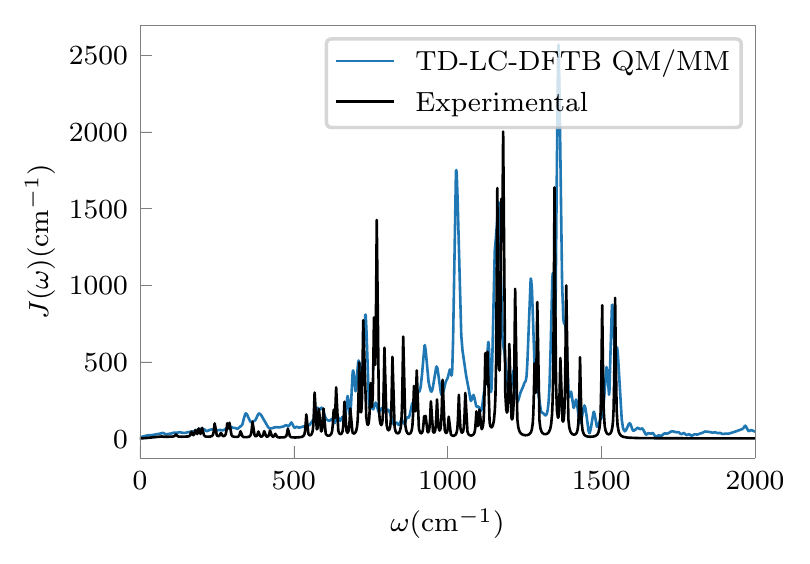}}
    \caption{Comparison of spectral densities averaged across all the eight BChl molecules obtained from~\citet{maityDFTBMMMolecular2020}}\label{fig:avg_spectral_densities}
\end{figure}

For the purposes of this exploration, we use the spectral densities calculated
via QM/MM trajectories. The two cases considered correspond to energy-gap
autocorrelation functions calculated using the TD-LC-DFTB method (average
reorganization energy of $\lambda = \SI{572.976}{\per\cm}$) and the ZINDO/S-CIS
semi-empirical method (average reorganization energy of $\lambda =
    \SI{839.032}{\per\cm}$) reported by~\citet{maityDFTBMMMolecular2020} In this
study, the site energies for the system Hamiltonian were calculated using
the same QM/MM MD trajectories and the off-diagonal terms were calculated
using TrESP~\cite{maityDFTBMMMolecular2020}. Comparison of the dynamics
under these recent spectral densities with that under the influence of a
spectral density calculated using classical trajectories and
ZINDO/S-CIS~\cite{olbrichTheorySimulationEnvironmental2011} is also shown.
In many of the later studies, the spectral densities were resolved for each BChl
unit and for each of the FMO monomers, effectively leading to 24 spectral
densities. However, for simplicity we are using the BChl site-specific spectral
densities averaged across the three monomers in FMO.  As discussed, the optical
excitation energies include influence from the solvents.  Consequently, the
electronic excitations, $\epsilon_j$, in the system Hamiltonian, $\hat{H}_0$,
are obtained from the site energies in Ref.~\autocite{maityDFTBMMMolecular2020}
by subtracting the site-specific reorganization energy. We used the correct
system Hamiltonian corresponding to the QM/MM MD spectral densities.

The various spectral densities, averaged across all the BChls, are plotted in
Fig.~\ref{fig:avg_spectral_densities}. We also show a spectral density obtained
using classical MD trajectories on ZINDO. The classical MD simulation of the
spectral density suffers from a significant blue shift of the high frequency
vibrations. Both the QM/MM spectral densities alleviate this problem.  Though
the alignment of the peaks is not perfect even when using QM/MM MD simulations,
it is quite close to the experimental fluorescence line narrowing spectral
density~\autocite{ratsepElectronPhononVibronic2007,
    kellShapePhononSpectral2013}. (Notice that the spectral density obtained using
classical MD and ZINDO is far smoother than the more recent spectral densities.
This is a result of a fitting procedure that was used
by~\citet{olbrichTheorySimulationEnvironmental2011}.)

There are a few questions that we want to focus our exploration on: (1)~How
important is choice of the density functional for characterizing the vibrations
and the protein scaffolding vis-{\`a}-vis the EET dynamics? (2) What is the
impact of the spatial inhomogeniety on the EET dynamics? (3)~How does one
analyze the individual pathways and routes of energy flow using exact dynamics
(4)~Given that BChl 3 is the sink of the EET process, how much of the excitation
energy goes into BChl 3 and how quickly? The last two questions, and their
connection with the features of spectral densities, is especially important in
terms of efficiency of the EET process.

\begin{figure}
    \centering
    \subfloat[TD-LC-DFTB]{\includegraphics{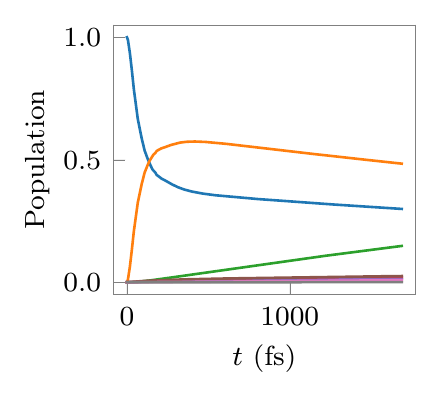}}
    \subfloat[TD-LC-DFTB Average Bath]{\includegraphics{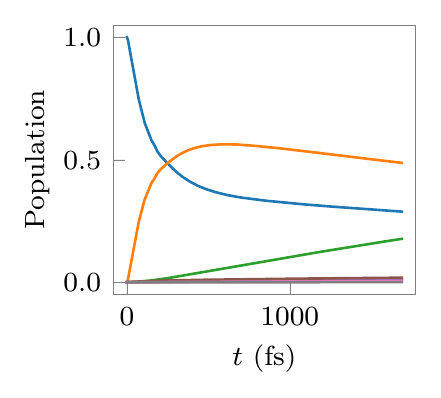}}
    
    \subfloat[ZINDO]{\includegraphics{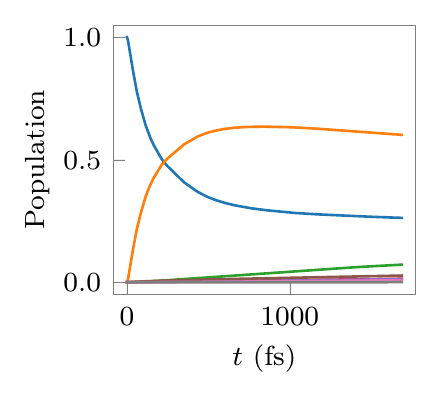}}
    \subfloat[ZINDO Average Bath]{\includegraphics{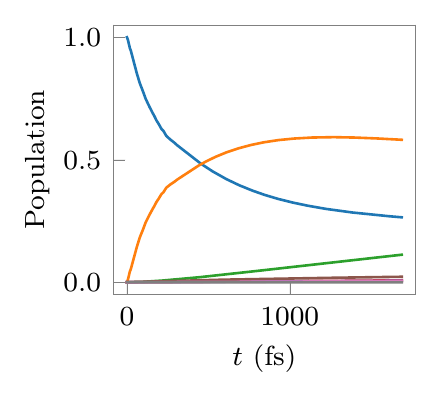}}
    \caption{Comparison of dynamics starting from $\tilde\rho(0)=\dyad{1}$ for the different \emph{ab initio} methods and with or without the spatial inhomogeneities. Blue: BChl 1. Orange: BChl 2. Green: BChl 3. Red: BChl 4. Purple: BChl 5. Brown: BChl 6. Pink: BChl 7. Gray: BChl 8.}\label{fig:dynamics1}
\end{figure}

\begin{figure}
    \centering
    \subfloat[TD-LC-DFTB]{\includegraphics{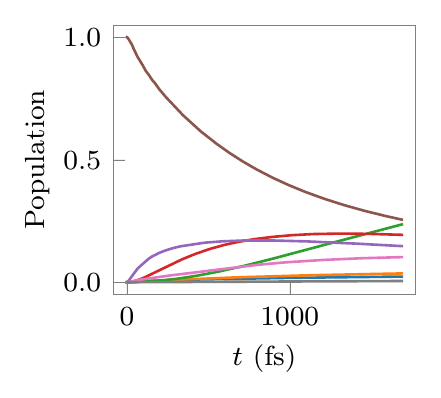}}
    \subfloat[TD-LC-DFTB Average Bath]{\includegraphics{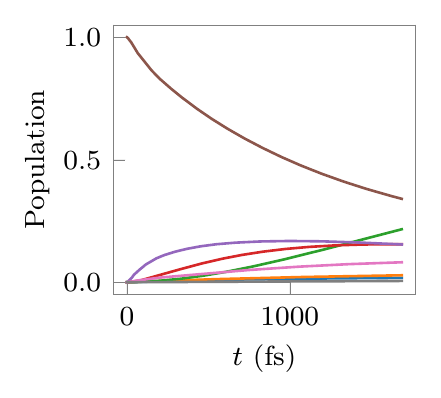}}
    
    \subfloat[ZINDO]{\includegraphics{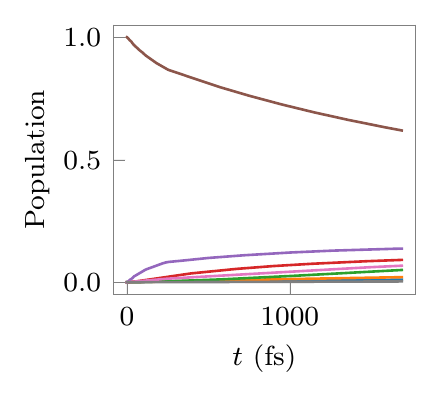}}
    \subfloat[ZINDO Average Bath]{\includegraphics{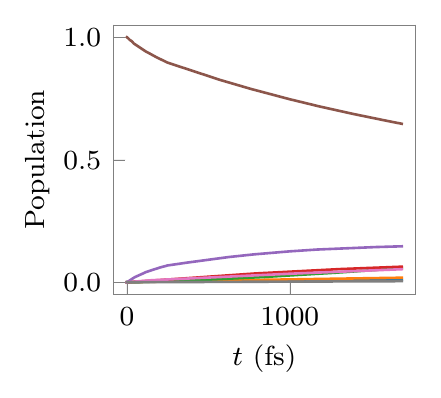}}
    \caption{Comparison of dynamics starting from $\tilde\rho(0)=\dyad{6}$ for the different \emph{ab initio} methods and with or without the spatial inhomogeneities. Colors same as Fig.~\ref{fig:dynamics1}.}\label{fig:dynamics6}
\end{figure}

First, let us consider the population dynamics corresponding to an initial
excitation of a single BChl unit. In Fig.~\ref{fig:dynamics1} and
Fig.~\ref{fig:dynamics6}, we show the dynamics that happen from each of the
different QM/MM MD spectral densities at an ambient temperature of
$T=\SI{300}{\kelvin}$ with initial conditions $\tilde\rho(0)=\dyad{1}$ and
$\tilde\rho(0)=\dyad{6}$ respectively. Figures~\ref{fig:dynamics1}
and~\ref{fig:dynamics6}~(a) and (c) correspond to the site specific spectral
densities, whereas Figs.~\ref{fig:dynamics1} and~\ref{fig:dynamics6}~(b) and (d)
correspond to the dynamics happening in the presence of the average
environments. It is interesting to note that while for both the methods, the
transfer into site 3 is faster in presence of the average spectral density when
$\tilde\rho(0)=\dyad{1}$, this is not the case when $\tilde\rho(0)=\dyad{6}$.
This seems to indicate that possibly not all evolutionary modifications to these
systems are geared towards an increased transport into the sink site 3. It is
instructive to note that it is not possible to come up with a very simple
conclusion about the effect of the spatial inhomogeneity, which seems to be
initial condition dependent.

The second aspect to consider from the Figs.~\ref{fig:dynamics1}
and~\ref{fig:dynamics6} is the effect of changing the method of calculating the energy-gap. The
most prominent difference between the spectral densities corresponding to the
two approaches is the higher intensity and consequently a larger
reorganization energy of the ZINDO spectral
density~\autocite{maityDFTBMMMolecular2020,
    chandrasekaranInfluenceForceFields2015}. This leads to a slowing down of
dynamics under ZINDO. It is noteworthy that despite this slowdown of the
dynamics, the transfer from site 1 to site 2 in Fig.~\ref{fig:dynamics1} is
increased from the TD-LC-DFTB functional to the ZINDO semi-empirical method.
Also, note that the bath inhomogenieties bring about a smaller difference in the
dynamics compared to the effect of changing from TD-LC-DFTB to ZINDO.

\begin{figure}
    \centering
    \subfloat[QM/MM $\tilde\rho(0) = \dyad{1}$]{\includegraphics{fig4c.pdf}}
    \subfloat[Classical $\tilde\rho(0) = \dyad{1}$]{\includegraphics{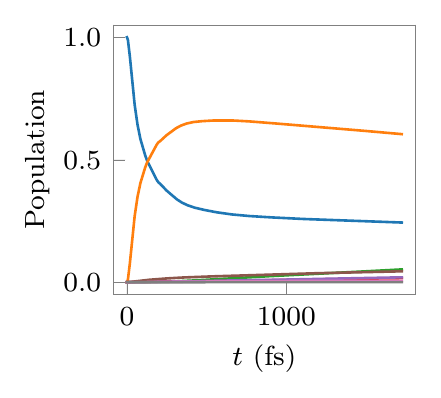}}
    
    \subfloat[QM/MM $\tilde\rho(0) = \dyad{6}$]{\includegraphics{fig5c.pdf}}
    \subfloat[Classical $\tilde\rho(0) = \dyad{6}$]{\includegraphics{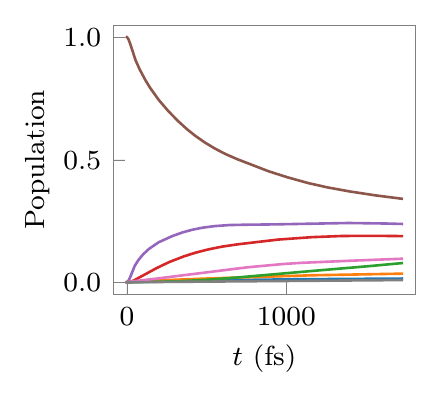}}
    \caption{Comparison of dynamics corresponding to ZINDO QM/MM MD spectral density versus ZINDO classical MD spectral density. Colors same as in Fig.~\ref{fig:dynamics1}.}\label{fig:class_dynamics_comparison}
\end{figure}

While it is known that resorting to classical MD trajectory-based simulations
generally leads to geometry mismatch problems, the actual impact of these errors
on the dynamics in this case has not been evaluated. The spectral density of the
classical ZINDO bath is shown in Fig.~\ref{fig:avg_spectral_densities}~(a). The
resultant spectral densities are clearly different from the QM/MM ones. However,
the effect of these differences on the dynamics is far from obvious. In
Fig.~\ref{fig:class_dynamics_comparison}, we show the dynamics corresponding to
the system coupled to site-local baths described by the classical ZINDO spectral
density. The system Hamiltonian corresponding to the ZINDO QM/MM MD simulation
is used with the classical spectral density as well. This has been done to
ensure that the effects we see are only coming from the spectral density. In
fact it is surprising, that despite the enormity of the differences at the
spectral density level, the dynamics is relatively similar to the QM/MM
calculation. The differences are quite subtle. In fact, it is arguable whether
the differences in dynamics caused by doing a qualitatively incorrect
calculation is greater than the ones seen by using a different method of
estimating the energy gap.

\begin{figure}
    \centering
    \subfloat[$\tilde\rho(0)=\dyad{1}$]{\includegraphics[scale=0.75]{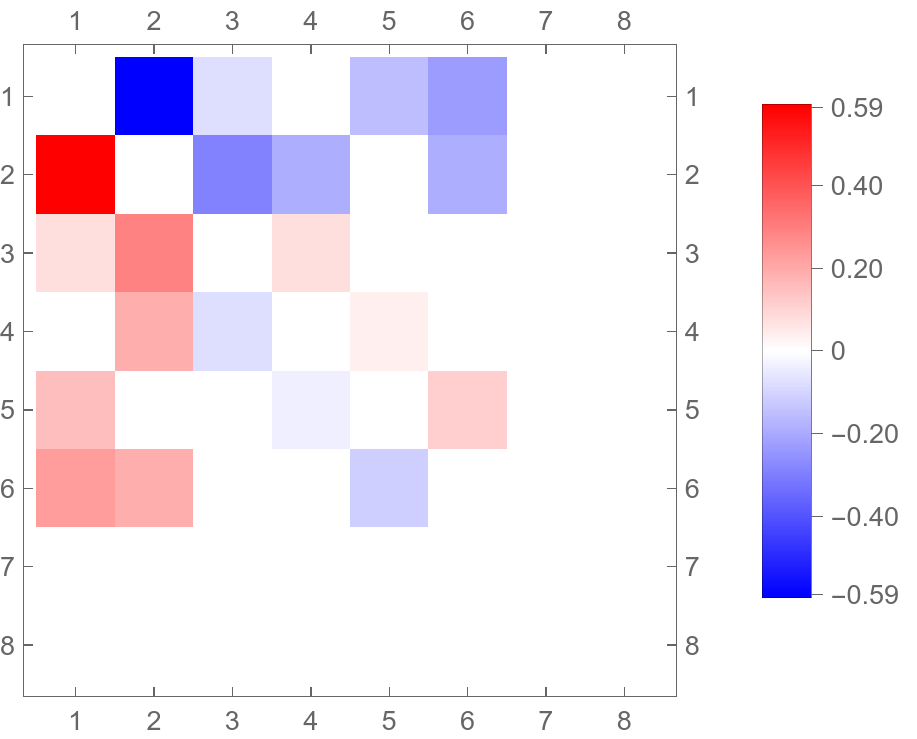}}
    
    \subfloat[$\tilde\rho(0)=\dyad{6}$]{\includegraphics[scale=0.75]{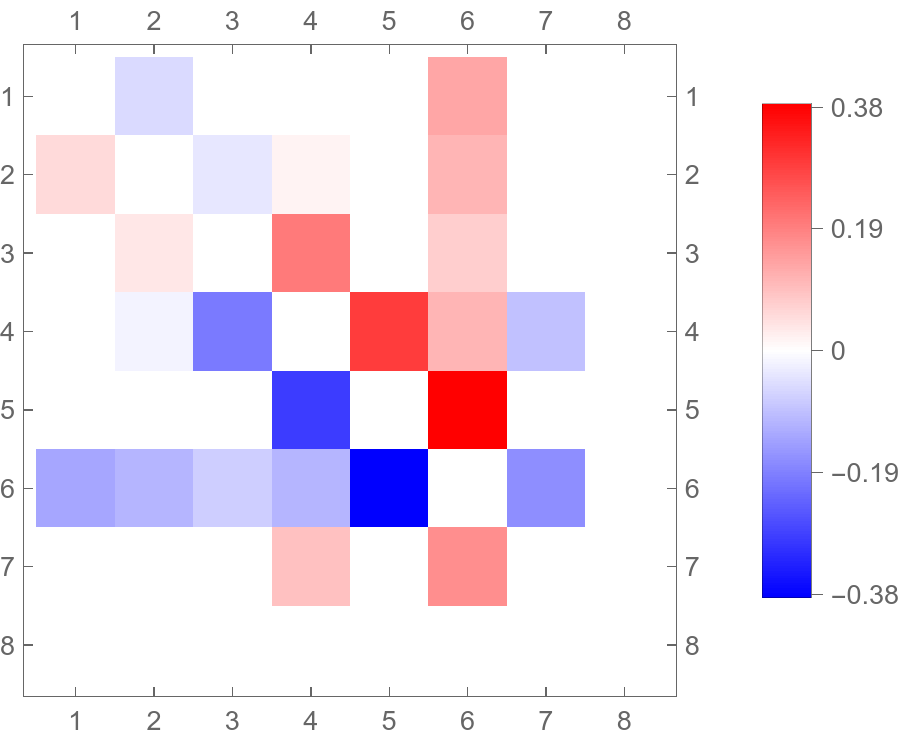}}
    \caption{State-to-state population transfer in presence of the TD-LC-DFTB baths for different initial conditions. The ($k$, $j$)\textsuperscript{th} cell in the image represents the average population transferred directly from site $j$ to site $k$, $\bar{P}_{k\leftarrow j}$.}\label{fig:full-channels}
\end{figure}

The effects of the method behind the excited state calculation and the spatial
inhomogeneity has been demonstrated on the EET. It is interesting to probe more
deeply into the mechanism behind these effects. Such a probe will reveal subtler
features of the dynamics than what was seen through the population dynamics.
Crucial to such an exploration is an understanding of the pathway-dependent
population transfer for each of the cases. \citet{wuEfficientEnergyTransfer2012}
have built flux networks to analyze the flows along different pathways. More
recently, Dani and Makri have shown that the time-derivative of the on-site
populations is related to linear combinations of the off-diagonal terms of the
reduced density matrix~\cite{daniQuantumStatetoStateRates2022}, and visualized
the coherences in forms of maps that encode information about the dynamics of the system~\cite{daniTimeEvolvingQuantumSuperpositions2022}. We
extended these ideas to partition the time-dependent population change on a site
in terms of transport along different state-to-state channels, which is useful
for understanding the instantaneous population transfer in these systems with
complex interconnects~\cite{boseImpactSolventStatetostate2023}. To explore the
overall importance of the different pathways using this coherence-based
analysis, we use a modified version of the integrated flux
approach~\cite{wuEfficientEnergyTransfer2012} and define a time-averaged
state-to-state population transfer:
\begin{align}
    \bar{P}_{k\leftarrow j} & = \frac{1}{T}\int_0^T\dd{t} P_{k\leftarrow j}(t),
\end{align}
where $P_{k\leftarrow j}(t) = -\frac{2}{\hbar}\mel{k}{\hat{H}_0}{j} \int_0^t
    \dd{t'} \Im\mel{k}{\tilde\rho(t')}{j}$ is the population transferred directly from
the $j$th site to the $k$th at time $t$. Here the time of integration is taken
to be $T\approx\SI{1500}{\fs}$. The main benefit of using a time-average instead
of a simple integration of the flux is that using the average one can capture
the relative speeds of transfer as well.

The complete information about the average population transfer along
each of these channels for the different initial conditions is presented in
Fig.~\ref{fig:full-channels}. This data corresponds to the TD-LC-DFTB spectral
density though the main trends carry over to the ZINDO spectral density as well.
Notice that for the case when $\tilde\rho(0) = \dyad{1}$, the most important
pathway is clearly $1\rightarrow 2\rightarrow 3$, with a secondary component
coming from the $1\rightarrow 6\rightarrow 5\rightarrow 4\rightarrow 3$. This is
known in the literature.  Non-insignificant contributions also happen along
$1\rightarrow 5\rightarrow 4\rightarrow 3$ and $1\rightarrow 2\rightarrow
    4\rightarrow 3$, both short-circuiting site 6. On the other hand, if
$\tilde\rho(0) = \dyad{6}$, the principle pathway is $6\rightarrow 5\rightarrow
    4\rightarrow 3$, with very tiny proportions of $6\rightarrow 1\rightarrow
    2\rightarrow 3$ and $6\rightarrow 2\rightarrow 3$.

\begin{figure}
    \centering
    \includegraphics[scale=0.4]{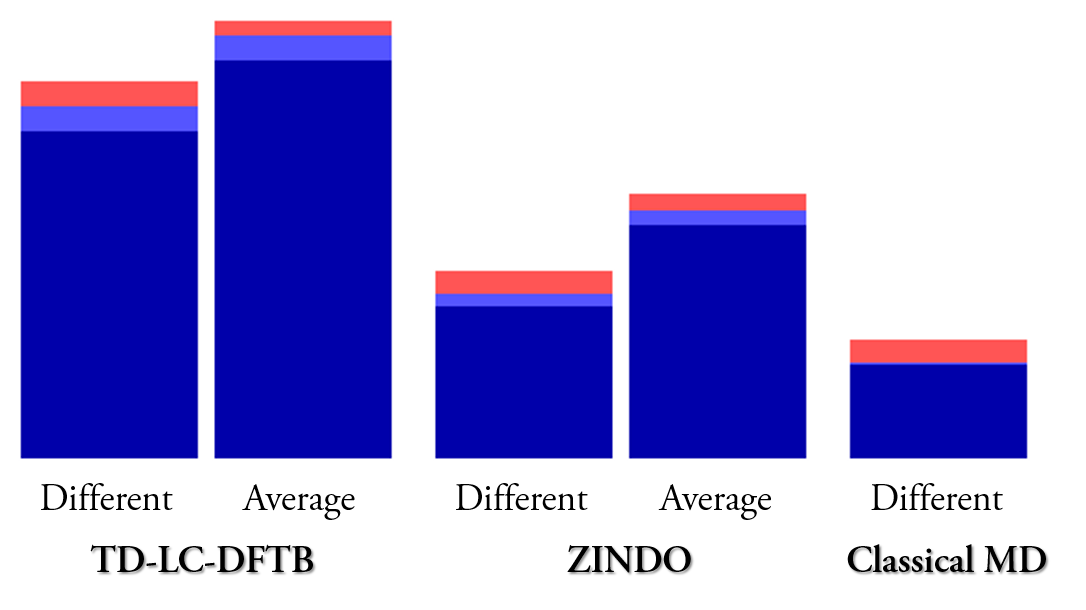}
    \caption{Time integrated transport to the 3\textsuperscript{rd} site,
        $\bar{P}^{(3)}$, when $\tilde\rho(0) = \dyad{1}$. Dark blue:
        $\bar{P}_{3\leftarrow 2}$. Light blue: $\bar{P}_{3\leftarrow 1}$. Red:
        $\langle\text{rest}\rangle$.}\label{fig:cdpt1}
\end{figure}

Now let us concentrate on contributions of the most important
pathways to the sink site 3. For our analysis, we divide the
population transfer from 1 to 3 into three components: (1) direct transport
from 1 to 3 ($\bar{P}_{3\leftarrow 1}$); (2) transport via 2 to 3
($\bar{P}_{3\leftarrow 2}$); (3) transport from all other sites combined which
encodes the possibility of transport in the $1\rightarrow 6\rightarrow
    5\rightarrow 4\rightarrow 3$ (represented as $\langle\text{rest}\rangle$). The
contributions of these three pathways into site 3 are shown in
Fig.~\ref{fig:cdpt1}.

\begin{figure}
    \centering
    \includegraphics[scale=0.4]{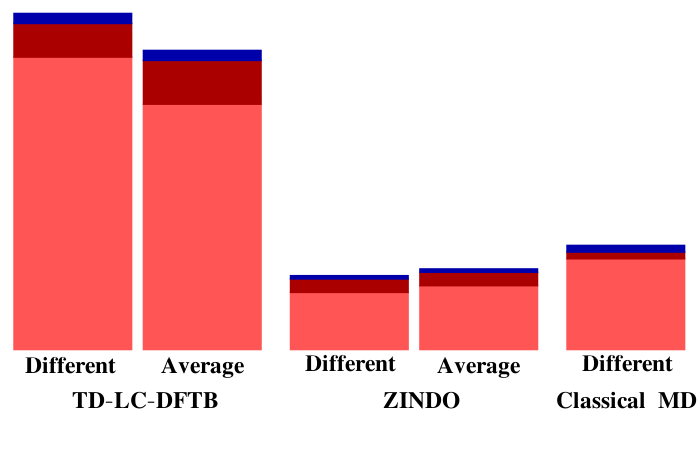}
    \caption{Time integrated transport to the 3\textsuperscript{rd} site,
        $\bar{P}^{(3)}$, when $\tilde\rho(0)=\dyad{6}$. Dark blue:
        $\bar{P}_{3\leftarrow 2}$. Dark red: $\bar{P}_{3\leftarrow 6}$. Red:
        $\bar{P}_{3\leftarrow 4}$.}\label{fig:cdpt6}
\end{figure}

The classical MD spectral density completely wipes out the direct transfer of
excitation from site 1 to 3. Though this direct route, $\bar{P}_{3\leftarrow
        1}$, is not exceptionally important, for the newer QM/MM-based spectral
densities, it gives a non-negligible contribution to the total population of
site 3. Also, note that the biggest impact of the spatial inhomogeneties is on
the $1\rightarrow 2\rightarrow 3$ pathway. More specifically, the transfer from
2 to 3 gets substantially inhibited by the presence of different site-specific
spectral densities. It is this inhibition of $\bar{P}_{3\leftarrow 2}$ that is
primarily responsible for the decrease of the overall transport into site 3.

In order to further explore the effect of the spatially inhomogeneous spectral
densities, let's consider in a bit more detail the percentage contribution of
the different components to the total transport. We have already mentioned the
overall increase in the transport to the 3rd site for the average spectral
densities. We notice that another effect that is consistently reproduced is the
fact the percentage contribution of the channel $\bar{P}_{3\leftarrow 2}$
consistently goes up from the site-dependent bath to the average bath. (The
percentage contribution of $\bar{P}_{3\leftarrow 2}$ goes up from approximately
$87\%$ to $90\%$ for TD-LC-DFTB and from approximately $81\%$ to $88\%$ for
ZINDO.) In contrast, the proportion of population along
$\langle\text{rest}\rangle$ is decreased quite significantly by the average
bath(close to $7\%$ to $3\%$ for the TD-LC-DFTB spectral density and from around
$13\%$ to $6\%$ for ZINDO). The direct transport along $\bar{P}_{3\leftarrow 1}$
also decreases in the presence of the average spectral density, but not by as
much a margin. (While here we have been talking in terms of percentage
contributions, the story is slightly different when it comes to the absolute
transfers along these channels. While $\bar{P}_{3\leftarrow 2}$ and
$\bar{P}_{3\leftarrow 1}$ both increase, the former much more significantly than
the latter, the absolute contribution along $\langle\text{rest}\rangle$ actually
decreases.) These subtle features would have been inaccessible in absence of a
method to probe the pathways of EET.

A similar analysis can be done for an initial excitation on the 6th monomer,
$\tilde\rho(0)=\dyad{6}$. Notice that in Fig.~\ref{fig:cdpt6}, the patterns are
even more convoluted than before. The most obvious thing that one can report is
that the major contributor to site 3 is site 4. Consulting
Fig.~\ref{fig:full-channels}, we realize that this is must be coming from the
major pathway of $6\rightarrow 5\rightarrow 4\rightarrow 3$. Contributions of
$6\rightarrow 4\rightarrow 3$ is minimal. The direct transport from site 2 to
site 3 is negligible in all cases. The conclusions with any degree of
universality stop there. Transport from 6 directly to 3 may be an important
factor, however, its importance seems to be dependent on the particular excited
state method used. Whereas the average spectral density gives a lesser transfer
into site 3 for the TD-LC-DFTB spectral densities, the reverse is true for
ZINDO. Additionally notice that the magnitude of differences between the
TD-LC-DFTB and the ZINDO bars in Figs.~\ref{fig:cdpt1} and~\ref{fig:cdpt6} is
larger than that between the ZINDO and the classical MD bars. Thus, the change
between the spectral densities is probably significantly less important than
that of changing the electronic excitation energies in the system Hamiltonian.
This once again demonstrates the growing need for accurate parameters for these
complex systems.

In this paper, we have evaluated the exact dynamics of the EET process in the FMO
complex when the ro-vibrational modes of the molecules and the protein scaffold
are described by accurate \textit{ab initio} molecular dynamics computations.
We have tried to shed light on the effects of the inhomogeneities in the
solvent by comparing with the dynamics corresponding to the average bath.
Surprisingly, the effects of changing the excited state method seem to be
possibly of a larger magnitude to that of removing spatial inhomogeneities or
even that of changing from QM/MM simulations to classical MD simulations.
Therefore, the noticeable differences in the spectral density must get washed
out when it comes to the dynamics. Additionally, this seems to imply that
estimating the correct electronic excitation energies of the BChl monomers is more
important than simulating an accurate spectral density.

There is a limit to the amount of information that can be extracted from the
direct population dynamics. We have used the recently introduced ideas of
relating the coherence to the population
transport~\cite{daniQuantumStatetoStateRates2022,
    daniTimeEvolvingQuantumSuperpositions2022,
    boseImpactSolventStatetostate2023} to explore the routes of transfer
that the molecular excitation takes. Given the very complex interactions between
these chromophores, it is not trivial to use an exact numerical computation to
attribute the excitation transport to the different directed path-ways that
exist. An analysis of the coherences is uniquely capable of answering these
questions, and we have used this technique to present a static picture of the
transport as it happens. In doing so, we have uncovered a couple of previously
unnoticed transport pathways for the FMO. An analysis of the pathway specific
contributions to the transport problem has yielded further evidence to support
the greater importance of the system Hamiltonian and electronic excitation
energies as opposed to the spectral density. These numerically ``exact''
approaches, in combination with better descriptions of the vibronic
interactions, promise to provide in-depth understandings of transport in similar
complex open systems.

\section*{Acknowledgment}
We thank Prof. Ulrich Kleinekath{\"o}fer and Sayan Maity for sharing their
spectral densities with us and for helpful discussions. AB acknowledges the
support of Princeton University and the Computational Chemical Science Center:
Chemistry in Solution and at Interfaces funded by the U.S. Department of Energy,
under Award No. DESC0019394 for providing resources for the simulation of some
of the dynamics.

% \appendix
% \section{Obtaining System Hamiltonian}\label{app:sysHam}
% is
% related to the electronic excitation energy of the site in
% the absence of the
% environment, $\epsilon_j$, as $E_j = \epsilon_j+\lambda_j$,
% where $\lambda_j=(1/\pi)\int_0^\infty \dd{\omega} J(\omega)/\omega$ is the
% reorganization energy for the $j$th bath. This is required to make the system
% Hamiltonian truly independent of the solvent effects, which is also an implicit
% assumption in the form of Eq.~\ref{eq:sys_bath} where the ``counter-term'' of
% reorganization energy is included with the bath.
% 

\bibliography{library}
% \bibliography{bibexport}
\end{document}